\documentstyle[aps,twocolumn,prl]{revtex}
\begin{document} 

\draft

% \preprint{ZHONG Fan et al}

\twocolumn[\hsize\textwidth\columnwidth\hsize\csname@twocolumnfalse\endcsname

\title
{Theory of Coupled Phase Transitions: Phase Separation and Abnormal Variation of Order Parameter}
\author{Fan Zhong$^{1,2}$,  M. Jiang$^1$, D. Y. Xing$^1$, and Jinming Dong$^1$}

\address{$^1$Department of Physics, Nanjing University, Nanjing 210093, P. R. China\\
$^2$Department of Physics, University of Hong Kong, Hong Kong, P. R. China}

\date{July 8, 1998}

\maketitle

\begin{abstract}
A simplified Ginzburg-Landau theory is presented to study generally a coupling of a first-order phase transition (FOPT) to a second-order phase transition (SOPT). We show analytically that, due to the coupling between the two phase transitions, the SOPT may exhibit a FOPT-like phase separation in which an ordered phase is separated from a disordered one. This phase separation results in a distinct behavior in the variation of the order parameter of the SOPT, namely, it is primarily the proportion of the ordered phase that contributes to the total order of the whole system during the FOPT. This growth mode may turn a mean-field critical exponent from 1/2 to 1 or even bigger.
\end{abstract}


]

% \narrowtext

%\section{Introduction} 
Phase transitions have been one of the fascinating fields in statistical and condensed matter physics, in which a recent theme is the behavior of transition metal oxides akin to the cuprate superconductor, a topic which fills most part of recent journals in association with such rich phenomena as charge and spin orders, stripe phases and phase separations. In complex systems, various phase transitions have been detected experimentally to coexist such as diffusion and shear~\cite{ko}, superconductivity and ferromagnetism~\cite{sf}, and antiferromagnetism and ferroelectrics~\cite{afe}. Meanwhile, electronic or magnetic transitions may compete with structural changes~\cite{sm}. For systems containing polymers like monolayers and liquid crystals, positional and orientational orders frequently come into play~\cite{ml}. As more sophisticated experimental techniques become available, more complex phenomena will certainly be revealed with more details, a theoretical understanding of such complex phenomena within the theory of phase transitions is therefore highly demanding.

In a recent Letter~\cite{zhong}, we put forward a Ginzburg-Landau theory based explicitly on the concept of coupled phase transitions. Taking the bainitic transformation as an example, we showed numerically that the coupling of two FOPTs representing respectively the diffusional and displacive transitions in a carbon-ion system leads to such possible phenomena as diffusional-phase-transition-induced nucleation of the displacive transition and incomplete reaction. Here, employing a simplified free-energy functional, we are able to show analytically that for a coupling of a FOPT with a SOPT, the latter transition may display a phase separation, a characteristic of a FOPT. This phase separation results in a distinct abnormal variation of the order parameter. Plausible arguments indicate that this in turn may lead to a critical exponent of 1 or even bigger instead of the normal mean-field value 1/2.

To be specific, we shall speak of a monolayer-like system composed of rodlike polymer molecules, each of which possesses an orientation such that there is a continuous orientational transition at a certain pressure $P$. A rough feature of the system is that as it is compressed, the order parameter $S$ of the orientation increases. In addition to this orientation transition, however, the molecules can change from a gaseous to a liquid state with compression. As a result, it is reasonable to assume that only in the liquid phase can the orientations of the molecules order, since the comparative small average distance of the molecules in the liquid renders their interactions more effectively. Therefore, our model system should be described by the coupling of the first-order gas-liquid and the second-order orientation phase transitions. One should keep in mind that such a concrete picture is only for the sake of convenience: the theory should be generally applicable. Thus, the density jump of the gas-liquid transition may correspond, for example, to a transition from an expanded to a condensed liquid states~\cite{lelc}.

Consider first the pure gas-liquid transition. By taking the order parameter $C({\bf x})$ as the area of molecules of one mole at the positions around ${\bf x}$, the (Gibbs) free-energy density for the gas-liquid transition is approximated by two pieces of parabolas around the gas and the liquid molar area $C_g^0$ and $C_l^0$, respectively, as
\begin{mathletters}
\label{fc}
\begin{eqnarray}
f_C(\{C({\bf x})\})=\frac{f_0}{2}\kappa \left(\nabla C({\bf x}) \right)^2+\left\{
\begin{array}{l}
\protect{f}_{g}^{0}, \ {\rm for} \  C>C_c,\\
f_{l}^{0}, \ {\rm for} \  C \leq C_c,
\end{array}
\right.
\end{eqnarray}
with $C_c$ being determined by continuity: $f_g^0(C_c)=f_l^0(C_c)$, which gives the barrier between the two phases, and
\begin{eqnarray}
f_{g}^{0}(\{C({\bf x})\}) &=& \frac{f_{0}}{2} \left(C({\bf x}) - C_g^0 \right)^{2}+(P-P_0)C({\bf x}),\nonumber\\
f_{l}^{0}(\{C({\bf x})\}) &=& \frac{f_{0}}{2} \left(C({\bf x}) - C_l^0\right)^{2}+(P-P_0)C({\bf x}),\label{fc0}
\end{eqnarray}
\end{mathletters}
where $P_0$ the equilibrium transition pressure. We have further simplified the free-energy by choosing identical positive coefficients $f_0$ and $\kappa$ for the two phases. From Eqs.~(\ref{fc0}), $C_c=(C_g^0+C_l^0)/2$, independent of $P$.

The special forms of the free-energy can be understood by their equations of states. For a uniform system at equilibrium, minimizing Eqs.~(\ref{fc}) without the gradient term, one obtains
\begin{eqnarray}
C_g(P) & = & C_g^0-(P-P_0)/f_0,\nonumber\\
C_l(P) & = & C_l^0-(P-P_0)/f_0.\label{ceq}
\end{eqnarray}
Consequently, $C_{g}^{0}$ and $C_l^0$ are the equilibrium molar areas at $P=P_0$, and $C_{g}$ and $C_l$ those at $P$. In the extremal case at $P=0$, the free-energy of the gas $f_g^0$ is lower than that of the liquid by $P_0 \Delta C$, where $\Delta C \equiv C_g^0-C_l^0 $ is the molar area difference between the two phases. Accordingly, the gas is stable. As $P$ increases, the difference in free-energy between the two phases decreases. At $P=P_0$, the two phases possess equal energies and so can coexist. At this pressure, the gas transforms to the liquid as the system is further compressed. When $P>P_0$, the liquid is stable and the gas metastable. We presume here that nuclei of the liquid phase have already been formed possibly by heterogeneous nucleations and so we do not consider nucleations anymore below.

For a nonuniform system at $P=P_0$, we assume that there is only one straight interface dividing the system into two parts. Minimizing the total free-energy $\int f_C d{\bf x}$ and taking into account proper continuity and boundary conditions, one obtains a one-dimensional-like interface at $x=0$ separating the gas and the liquid as, 
\begin{eqnarray}
C(x \leq 0, y)&=&C_l^0+\frac{\Delta C}{2} \exp \left( x/ \sqrt{\kappa} \right), \nonumber\\
C(x \geq 0, y)&=&C_g^0-\frac{\Delta C}{2} \exp \left( -x/ \sqrt{\kappa} \right). \label{cif}
\end{eqnarray}
Note that Eqs.~(\ref{cif}) yield correct equilibrium molar areas at $P=P_0$ for both phases (Eqs.~(\ref{ceq})) beyond the interfacial region of width $2 \sqrt{\kappa}$. This is obtained at the expense of maintaining different internal pressures $P_0-f_0 \sqrt{\kappa} \Delta C/2$ and $P_0+f_0 \sqrt{\kappa}\Delta C/2$ for the liquid and the gas, respectively, an artifact stemming from the discontinuous jump in the slopes of the free-energy, Eqs.~(\ref{fc}), at $C=C_c$. This jump exerts opposite pressures of the magnitude $f_0 \sqrt{\kappa} \Delta C/2$ on both sides of the interface such that the total external pressure of the system is again $P_0$, the transition point. However, the artifact will not affect our following analyses; we only need a robust linear interface separating two phases with their respective states given by the equilibrium equations, Eqs.~(\ref{ceq}). 

Next let us turn to the orientation phase transition. By defining a local coarse-grained order parameter $S({\bf x})$ as the average orientation per unit area around ${\bf x}$, the free-energy density for this part alone is assumed to be
\begin{equation}
f_S(\{S({\bf x})\})=\frac{1}{2}r S^{2}({\bf x})+\frac{1}{4}u S^{4}({\bf x}) + \frac{1}{2} \kappa' \left( \nabla S({\bf x}) \right)^2,\label{fs}
\end{equation}
where $r=a(P_0-P)$ and $a$, $u$, and $\kappa'$ are positive constants. We have assumed that $S$ possesses inversion symmetry. Eq.~(\ref{fs}) represents a pressure-induced continuous phase transition at $P_0$ from a disordered phase with $S=0$ to an ordered phase with $S=\sqrt{a(P-P_0)/u}$ as $P$ increases. This gives a critical exponent $\beta=1/2$ defined by $S \propto (P-P_0)^{\beta}$~\cite{cp}. If a system is compressed from an area $A_0$ at $P_0$ to $A$ at $P$ no matter whether it is in the pure gaseous state or in the pure liquid state, Eqs.~(\ref{ceq}) imply that $P-P_0 \propto A_0-A$~\cite{note}, since the total molar number of the system is conserved. Therefore the variation of $S$ with $A_0-A$ during compression is the standard mean-field law with $\beta=1/2$. However, these results may be radically changed due to the gas-liquid coexistence.

We now proceed to the coupling of the two phase transitions. Within the framework of Ginzburg-Landau theory, the simplest coupling complying with the symmetry requirement of both transitions is given by 
\begin{equation}
f_{CS}(\{C({\bf x})\},\{S({\bf x})\}) = \frac{1}{2} g \left(C({\bf x})-C_{0} \right) S^{2}({\bf x}),\label{fcs}
\end {equation}
where $g$ is the coupling coefficient, and $C_0$ a constant. It is expected that the smaller the molar area, the greater the orientational order. Taking stability into account, we have $0 < g <\sqrt{2uf_0}$ (see Eqs.~(\ref{cseq}) below). Since we have presumed that both transitions take place at the same $P_0$, $C_0$ may be regarded as an adjustment of the transition point. $f_{CS}$ plusing $f_C$ and $f_S$ constitutes the total free energy density $f(\{C\},\{S\})$ of the system.

The effect of the coupling is most clearly seen from the renormalized prefactor of $S^2({\bf x})$,
\begin{equation}
r'=r+g \left( C({\bf x})-C_0\right),\label{rp}
\end {equation}
which is position dependent. As $C$ takes different values in the gas and the liquid, the orientation phase transition can behave differently. Physically, it is clear that in the gaseous region, the molecules are far apart. As a result, interactions among their orientations are weak and so cannot afford order. In the liquid phase, on the other hand, they occupy such a small space that they must simultaneously order orientationally. Accordingly, {\em in the gas, $r' \geq 0$ and so $S=0$; while in the liquid, $r' < 0$ and so $S > 0$}. This is the most important {\it ansatz} whose consequences will be explored below.

The first result of our ansatz is that for a uniform system, the equilibrium solution now becomes
\begin{eqnarray}
&&C_g'(P)=C_g(P)=C_g^0-(P-P_0)/f_0,\nonumber\\
&&S_g (P) = 0,\nonumber\\
&&C_l'(P)= C_l(P)-\frac{g}{2f_0}S_l^2(P),\nonumber\\
&&S_l^2 (P) = B(P-P_l)/g,\label{cseq}
\end{eqnarray}
where $B=g(a+g/f_0)/(u-g^2/2f_0)$ is a dimensionless constant, and $P_l=P_0+\frac{g (C_l^0-C_0)}{a+g/f_0}$ is the pressure at which a pure liquid would begin to order orientationally. These expressions show that at a given pressure, the coupling makes the molar area of the liquid reduced because the molecules tilt up or straighten on ordering as expected, whereas that of the gas retains their disordered value. Meanwhile, there is a finite net orientation in the liquid. Using these expressions and Eq.~(\ref{rp}), the ansatz that the liquid is ordered orientationally and the gas disordered can be rephrased as a restriction on $P$ as
\begin{equation}
P_l < P \leq P_0+\frac{g (C_g^0-C_0)}{a+g/f_0} \equiv P_g, \label{ansatz}
\end{equation}
where $P_g$ is the correspondence of $P_l$ for the gas, {\it i.e.}, the pressure at which a pure orientationally disordered gas would start to order orientationally had it not transformed into a liquid. 

Now we determine the pressure where both phases possess an equal free-energy, {\it i.e.}, the transition pressure $P_t$, which must of course be bigger than $P_l$. To this end, it is noted that the orientational ordering of the liquid state lowers its free-energy as compared to orientational disordering. As a consequence, the relative depth of the two free-energy wells is so reduced that now a lower pressure than $P_0$ can eliminate the difference, since the gas remains orientationally disordered. This implies that we must have $P_l<P_t<P_0$. This is reasonable because otherwise if $P_l>P_0$, the system will turn into the liquid state before it becomes ordered. Keeping this in mind, one can get from Eqs.~(\ref{cseq}) and the total free-energy,
\begin{equation}
P_t = P_l + \frac{2(P_0-P_l)}{1+\sqrt{1+B(P_0-P_l)/ (P_g-P_l)}}.
\end{equation}
Note that the transition point of a FOPT is generally only the point at which both phases obtain equal free-energy. It is not a singularity point of the free-energy.

Having known $P_t$, one can determine the constraint on $C_0$ from the definitions of $P_l$ and $P_g$ and Eq.~(\ref{ansatz}), {\it i.e.}, the range where $P_t$ is available. First the physical outcome of $P_l<P_t<P_0$ implies $C_0>C_l^0$. Then, on the one hand, if $P_g \geq P_0$, or equivalently $C_0 \leq C_g^0$, meaning that a pure gas would start to order orientationally beyond $P_0$ had it not been transformed into liquid, one gets $P_l<P_t<P_0$ and so $C_l^0<C_0 \leq C_g^0$. On the other hand, if $P_g<P_0$ or $C_0>C_g^0$, then $P_t$ must be smaller than $P_g$. This leads to $C_0<C_g^0+B \Delta C/4$, and so $C_g^0<C_0<C_g^0+B \Delta C/4$. In this case, the coupling pushes $P_t$ to far below $P_0$. 

Here come our main results. As the system is compressed, the pressure increases. When $P$ reaches $P_l$, it would be possible for the liquid to start to order orientationally had it formed. Yet, such liquid embryos possess a high energy than the bulk gas and so are doomed to dissolve as $P$ is still smaller than $P_t$. Only at $P_t$ can they become stable to further grow (if they are larger than the critical nucleus). On the other hand, once the liquid nuclei are formed, they must simultaneously be orientationally ordered, otherwise they are still not stable enough to survive since $P$ is less than $P_0$, at which an orientationally disordered (but unstable with respect to orientational order) liquid acquires equal energy to a disordered gas. Consequently, only those orientationally ordered liquid regions remains and can live with the gas. The system therefore displays a physical picture in which an orientationally disordered gas coexists with an orientationally ordered liquid with an interface between them. The interesting point here is that the continuous orientation phase transition exhibits an inhomogeneous $S({\bf x})=0$ and $S({\bf x})>0$ phase separation, a characteristic of first-order phase transitions. 

As a result of this phase separation, the variation of $S$ demonstrates a distinct mode. As the system is further compressed, the pressure remains at $P_t$ like an ordinary gas-liquid transition. The total area of the system is reduced by transforming the gas into the liquid. In so doing, the new point here is that the random orientations of the molecules get gradually align as they pass the interfaces such that the order parameter acquires a constant local value $S_l(P_t)$ in the formed liquid. The physical reason behind is the interactions among the orientations effect progressively as their average distances reduce from the gas to the liquid and then remain at the respective constant local values in the two phases. Accordingly, as the amount of the liquid increases, the total or integrated orientation order of the system increases, but $S_l(P_t)$ remains constant. It is therefore the expansion of the total liquid regions with a constant $S_l(P_t)$ rather than the increase of the local $S({\bf x})$ itself that contributes to the increase of the total orientation order of the whole system during compression.

The above analyses are exact and independent of the spatial dimensionality $D$ (except Eqs.~(\ref{cif})). In the following we give plausible arguments to see how the abnormal mode of growth of the order parameter is different from the normal $1/2$-mean-field law.

At the beginning of the transition, all molecules in the system with an area $A_0$ are in the gaseous state with a molar area $C_{gt}'$ (the extra subscript $t$ stands for those values at $P_t$ hereafter). If the system is compressed to an area $A$ at the constant $P_t$, the orientationally ordered liquid is formed and separated from the orientationally disordered gas by interfaces, which is a resultant of the gas-liquid and the order-disorder interfaces. We assume below that the interfaces have such a thin width (so long as $\kappa$ and $\kappa'$ are sufficiently small) that they can be separated from the bulk of the liquid and so the mean-field approach is valid. Let $A_{i}$ denote the total area of the interfaces, $C_{i}$ and $S_{i}$ the average molar area and the average orientation per unit area in the interfacial region, respectively. By noting that the total number of molecules is conserved, the total area of the liquid state is then
\begin{equation}
A_{l}=\frac{C_{lt}'}{C_{gt}'-C_{lt}'}(A_0-A)-\frac{C_{lt}'}{C_{i}} \frac{C_{gt}'-C_{i}}{C_{gt}'-C_{lt}'}A_{i}, \label{al}
\end{equation}
a kind of the lever rule with the interface correction. Accordingly, the average orientational order parameter of the whole system is given by
\begin{equation}
\overline{S} = (S_{lt}A_{l}+S_{i}A_{i})/A.
\end{equation}
As pointed out above, within the usual mean-field theory of critical phenomena, $\overline{S} \propto (A_0-A)^\beta$ with $\beta=1/2$ independent of the spatial dimensions. We now argue how $\beta$ is changed by the distinct growth mode. 

If the system is bisected by a linear interface like Eqs.~(\ref{cif}), its effective dimensionality is 1. In this case, $A_{i}$ is a constant proportional to the width of the interface. Consequently, $\overline{S}$ is proportional to $A_0-A$, and so $\beta=1$. For a system of higher dimensions, circular or spherical interfaces render $A_{i}$ grow with $A_{l}$ (For $D>2$, $A$ means (hyper)volume). Assume that the width of the interfaces does not change during growth (accordingly, neither do $C_{i}$ and $S_{i}$). For a liquid region of radius $R$, $A_{i} \propto R^{D-1}$ and $A_{l} \propto R^D$. Accordingly, for not too small size of the liquid regions, the bulk contribution $A_{l}$ is dominant and so $\overline{S}$ is proportional to $A_0-A$ to the leading order, giving rise to $\beta=1$. For small size, on the other hand, $A_{i}$ is dominant. Accordingly, $A_{i} \propto A_0-A$ from Eq.~(\ref{al}) and so $A_l \propto (A_0-A)^{D/(D-1)}$ to the leading order. However, since the factor before $A_{i}$ in Eq.~(\ref{al}) is less than one, and $S_{lt}$ is larger than $S_{i}$, also the width of the interfaces may change for too small size, the two leading orders with the respective exponent $D/(D-1)$ and 1 of $\overline{S}$ vs $(A_0-A)$ may be comparable. Therefore, the critical exponent of the order parameter, $\beta$, may appear to be in the range of 1 and 2 (for $D=2$) instead of the normal 1/2. A result of this competition at the beginning of the transition (but still beyond the critical region) is that the variation of $\overline{S}$ vs $A_0-A$ may become {\em concave} rather than the normal {\em convex} of $\beta=1/2$. This result may be relevant to some experiments~\cite{exp}.

%\section{Conclusion}
In this paper, we have attempted to provide an overall theoretical understanding of the coupling of phase transitions in complex systems by studying a model system consisting of a first-order gas-liquid transition and a concomitant second-order orientational phase transition. Using a simplified Ginzburg-Landau theory, we are able to show analytically that coupling of the two transitions may make the continuous orientational phase transition exhibit a phase separation of an ordered state from a disordered state. In particular, as the system transforms at a constant pressure $P_t$ from gas to liquid, the random orientations of the molecules become ordered continuously as they pass through the interfaces between the gas and the liquid, and then acquire a constant local value $S_l(P_t)$ in the formed liquid. Such a phase separation gives rise to an abnormal growth law in which the order parameter of the orientation transition increases with the expansion of the ordered liquid regions during compression. This growth law in turn makes the critical exponent $\beta$ change from the 1/2-mean-field value to 1 or even bigger. 

The prerequisite of these results is based on the ansatz that only the liquid is orientationally ordered. This requires that the coupling lowers the transition pressure. Critical fluctuations should not change the overall picture once $P_t$ is not too close to $P_g$. Neither should higher order couplings since they only renormalize the higher order terms in the free-energy.

We emphasize that the theory presented should be applicable to other systems though we focus our attention only on a monolayer-like system consisting of polymers. A salient feature of the complex system under consideration is that it contains two types of order that can interplay. Another entity besides polymers that may possibly demonstrate the phenomena discussed here may be, for example, electrons, which possess charge and spin. When the electrons make up condensed matters as an important component, their position order such as charge density waves and even the superconducting condensate may well compete with their spin order via electromagnetic interactions. So one may readily imagine analogous pictures to be happened in such systems as well.

This work was supported by the Postdoctoral Science Foundation of China.

% \newpage


\begin{references}
\bibitem{note}In general, $\Delta P \propto \Delta V$ irrespective of the detailed form of the equation of states, provided that the compressitivity is constant.
\end{references}
\end{document}